\documentclass[%
 reprint,
 amsmath,amssymb,
 aps,
]{revtex4-2}

\usepackage{graphicx}
\usepackage{dcolumn}
\usepackage{bm}
\usepackage{mathtools}

\usepackage{accents}

\newcommand{\doublehat}[1]{{\cal \hat #1}}

\usepackage{xcolor}

\begin{document}

\preprint{APS/123-QED}

\title{Finite memory time and anisotropy effects for initial magnetic energy growth in random flow of conducting media}
\author{Illarionov E.A.}
\affiliation{
 Department of Mechanics and Mathematics, Moscow State University, Moscow, Russia
\\
Moscow Center of Fundamental and Applied Mathematics, Moscow, Russia}
\author{Sokoloff D.D.}
  \affiliation{Department of Physics, Moscow State University, Moscow, Russia \\
  Moscow Center of Fundamental and Applied Mathematics, Moscow, Russia}

\date{\today}

\begin{abstract}
The dynamo mechanism  is a process of magnetic field self-excitation in a moving
electrically conducting fluid. One of the most interesting applications of this mechanism related to the astrophysical systems is the case of a random motion of plasma. For the very first stage of
the process, the governing dynamo equation can be reduced to a system
of first order ordinary differential equations. For this case we
suggest a regular method to calculate the growth rate of magnetic
energy. Based on this method we calculate the growth rate for random
flow with finite memory time and anisotropic statistical
distribution of the stretching matrix and compare the results with
corresponding ones for isotropic case and for short-correlated
approximation.  We find that for moderate Strouhal numbers and moderate
anisotropy the analytical results reproduce
the numerically estimated growth rates reasonably well, while
for larger governing parameters the quantitative difference becomes
substantial.  In particular, analytical approximation is applicable for the Strouhal numbers $s<0.6$ and we find some
numerical models and observational examples for which this region might be relevant. Rather unexpectedly, we find that the mirror
asymmetry does not contribute to the growth rates obtained, although the
mirror asymmetry effects are known to be crucial for later stages of
dynamo action.    
\end{abstract}

\maketitle

\section{Introduction}

Magnetic field in various celestial bodies is believed to be excited by hydromagnetic dynamo, i.e. by the transformation of kinetic energy of electrically conducting medium to magnetic field energy due to electromagnetic induction (e.g. \cite{1978Moffatt, 2004Holl}). In last two decades the dynamo mechanism could be realized even in laboratory experiments (e.g. \cite{Fetal14}). Contemporary abilities of numerical simulations are sufficient to reproduce dynamo action at least for appropriate range of dynamo governing parameters, which are however sometimes quite remote from realistic ones for particular celestial bodies. Thus analytical or quasi-analytical (asymptotical) results  in dynamo studies  remain important for extrapolation of numerical results in the extended parametric domain as well as in case where our knowledge about details of inner structure of a celestial body is very limited (e.g. \cite{2012Br}). 

The bulk of analytical results accumulated in dynamo studies during the half a century of intensive investigations is obviously limited by various simplifications used. In particular, consideration of statistically isotropic and homogeneous random flows (convection and/or turbulence) and/or short-correlated (in time) approximation are usual limitations of the approach (e.g. \cite{1990Z}) and an intention to remove these limitations at least to some extent looks in our opinion attractive. A development in this direction is the aim of our paper.

To be specific, we focus our attention on a certain  stage of dynamo instability, which happens to be the very first stage of magnetic field evolution. Provided the seed magnetic field, which is involved in the dynamo action is a large-scale one, the Ohmic losses can be neglected at the initial stage of magnetic field evolution $t \ll t^*$. Using Lagrangian coordinates one can reduce the governing equation for magnetic field to an ordinary differential equation. We investigate only this particular stage and stress that the approach becomes irrelevant in later stages of magnetic field evolution. In addition, we suppose that the seed magnetic field is weak enough to be dynamically unimportant for $t<t^*$ as well as we admit that the memory time $\tau$ of velocity field is much smaller than $t^*$. Consideration of different combination of these timescales is obviously interesting, however, such an approach requires specific methods for elaboration.

One other point is that we investigate systematically just one quantity associated with dynamo, i.e. magnetic energy averaged over turbulent (convective) pulsations, and leave aside such important quantities as the mean magnetic field and etc., for further studies. This approach has its origins in the so-called Kazantsev--Kraichnan model \cite{1967KN, K68}, which is suggested for short-correlated statistically isotropic and homogeneous random flows.  For the initial stage of magnetic field evolution we develop the results for finite correlation time and statistically anisotropic random flows. 

We relate the results for the finite correlation time and statistically anisotropic flows to corresponding ones for short-correlated (in time) statistically isotropic flows to separate effects of memory and anisotropy. We would like to stress once again that our present research is adequate for only the initial stage of magnetic field evolution.

The rest of the paper is organized as follows. 
First, we describe how to calculate growth rate of
the second statistical moment of a vector field in the flow models with finite memory time $\tau$.
This method is not only
limited to magnetic fields and can actually be considered as a generic one. Then we consider different velocity
fields in 2D and 3D spaces (isotropic and anisotropic) and derive the growth rates for magnetic energy. In Appendix A we demonstrate that
an alternative approach, which is applicable to isotropic models only, provides the same
growth rate approximation. Finally, in Appendix B
we discuss the limiting case $\tau\to 0$ and its relation to the short-correlated approximation.

\section{Growth rate of higher statistical moments of a vector field}

In this section we describe a generic approach for the
calculation of growth rates of higher statistical moments of vector fields in the flow models with finite memory time. We will use the following notations:
\begin{enumerate}
    \item bold letters denote vectors (e.g. $\bf H$, $\bf x$)
    \item capital letters denote matrices (e.g. $A$, $B$)
    \item calligraphic capital letters denote extended matrices (will be defined later, e.g. $\cal A$,  $\cal E$)
    \item hat symbol denotes realization of a random process on time intervals between renovation instants (will be defined later, e.g. $\hat A_i$, ${\cal\hat A}_i$)
\end{enumerate}

Let ${\bf H} = (H_1, ..., H_k)^T$ be a $k$-dimensional vector field and its evolution is defined  by the equation
\begin{equation}
    \frac{d \bf H}{d t} =  A(t) {\bf H}\, 
\label{vector_eq}    
\end{equation}
with the initial condition ${\bf H}(0) = {\bf H}_0$
(all variables are unitless).
Note that the dimensionality given by  $k$ is not fixed in advance (e.g. $k=3$) because the model can be similarly generalized to higher dimensions. 

We assume that the $k\times k$ matrix denoted by $A(t)$ describes a piece-wise temporally constant random process. Now by specifying some arbitrary positive $\tau$ and considering the following
time instants $\tau$, $2 \tau$, $3\tau$, ... , which are called as the renovation instants, we have on each such time interval of the form $[n\tau, (n+1)\tau)$, the condition that the matrix $A(t)$ is independent of time and on other different intervals and therefore these matrices are statistically independent and identically distributed (i.i.d.). Such a model is called the renovation model.

Let $\hat A_i$ denotes a renovation instant of the matrix $A(t)$ on the $i$-th time interval. Solution of Eq.~(\ref{vector_eq}) at time instants $n\tau$ can then be expressed in terms of  the product of matrix exponentials:
\begin{equation}
    {\bf H}(n\tau) = \exp({\hat A}_n\tau) \exp( {\hat A}_{n-1}\tau)...\exp( {\hat A}_1\tau) {\bf H}_0\, .
\label{vector_sol}    
\end{equation}
Averaging Eq.~(\ref{vector_sol}) and using the i.i.d. property of matrices
$A_i$, we arrive at the following equation
\begin{equation}
    \langle{\bf H}\rangle(n\tau) = \langle \exp ( {\hat A}\tau) \rangle^n {\bf H}_0\, ,
\label{vector_avr}    
\end{equation}
where $\hat A$ denotes a random matrix with the same distribution as any of the matrices ${\hat A}_i$. Now the Eq.~(\ref{vector_avr}) implies that growth rate of any component of 
the mean vector field  $\langle{\bf H}\rangle$ is given by the
leading eigenvalue of the mean matrix exponential $ \langle  \exp ({\hat A}\tau) \rangle$.

Now we recall the definition of growth rates $\gamma_p$ for the higher statistical moment of a vector field:
\begin{equation}
    \gamma_p = 
    \lim\limits_{n\to\infty}\frac{1}{2pn\tau}\ln\langle  \|{\bf H}(n\tau)\|^p \rangle \, .
\end{equation}
Here an integer $p$ denotes the order of the statistical moment and $\tau$ is
the renewal time (which is twice the memory time). Growth rate of the mean energy therefore corresponds to $p=2$.

In order to obtain growth rate of higher statistical moments of the vector field, e.g. of the second moment
$\langle{\|\bf H}\|^2\rangle$,
it is desired to have a differential equation for $\|{\bf H}\|^2$. In some 
exceptional cases (e.g. if the distribution of matrices $A_i$ is isotropic)
one can omit this step as we discuss in Appendix A. However, in any case such a deviation from the isotropic one makes the method inapplicable. 

An equation for 
$\|{\bf H}\|^2$ (as well as for higher moments after appropriate modification) can be obtained as follows.
Consider a quantity with double index given by $Z_{ij} = H_i H_j$ and arrange the vector ${\bf Z} = (Z_{11}, Z_{12}, ..., Z_{kk})^T$ of length $k^2$.
Now applying the  Leibniz's rule for calculating $\partial Z_{ij} /\partial t = \partial (H_i H_j) /\partial t$ and Eq.~(\ref{vector_eq}),
we obtain an equation for ${\bf Z}$:
\begin{equation}
    \frac{d \bf Z}{d t} = {\cal A}(t) {\bf Z}\, .
\label{vector_ext}    
\end{equation}
Here ${\cal A}(t)$ is a $k^2\times k^2$ matrix and its elements
can be easily expressed in terms of the matrix $A(t)$ as follows:
\begin{equation}
   {\cal  A}_{k(i-1)+j, k(l-1)+m} = {A}_{il}\delta_{jm} + {A}_{jm}\delta_{il} \, .
   \label{A_hat}
\end{equation}
We refer to the Eq.~(\ref{vector_ext}) as the extended equation and use
calligraphic script (e.g. ${\cal  A}$, $\doublehat  A$) to denote the associated matrices and naturally name them as the extended matrices. Obviously, Eq.~(\ref{A_hat}) holds also for matrices with a hat symbol (i.e. realization of the process on a specific renovation interval). Repeating arguments used in 
Eq.~(\ref{vector_sol}) and (\ref{vector_avr}) we obtain that the growth rate of the mean vector $\langle \bf Z \rangle$ is defined by the leading
eigenvalue of the mean matrix exponential
$\langle\exp(\doublehat A \tau)\rangle$. It remains to
note that all of $\langle H_i^2 \rangle$ are within the components 
of the vector $\langle {\bf Z} \rangle$, thus the growth rate of
$\langle {\bf Z} \rangle$ gives the growth rate of $\langle{\|\bf H}\|^2\rangle$.

An exact relation between leading eigenvalue (denoted $\lambda$) 
of the mean  matrix exponential $\langle\exp(\doublehat A\tau)\rangle$ and $\gamma_2$ can be derived as 
follows. By definition of $\gamma_2$, it follows
that $\langle \|{\bf H}\|^2\rangle$ grows as $\exp(4\gamma_2n\tau)$ for large $n$.
At the same time, each $H_i^2=Z_{ii}$ grows as $\lambda^n$ for large $n$.
It follows that $\gamma_2 = (1/4\tau)\ln \lambda$.

Now we summarize the steps for calculation of the growth rate $\gamma_2$
in the context of the renovation model:
\begin{enumerate}
    \item Proceed from the initial Eq.~(\ref{vector_eq}) to the
    extended Eq.~(\ref{vector_ext});
    \item Find the leading eigenvalue $\lambda$ of the mean matrix exponential 
    $\langle \exp(\doublehat A \tau)\rangle$;
    \item Finally $\gamma_2 = (1/4\tau)\ln\lambda$.
\end{enumerate}

The described method for computation of the growth rate of the second
moment can be naturally extended to higher-order moments.
E.g. for the moment of order 3 one has to consider quantities with a triple index
$Z_{ijl}=H_i H_j H_l$. This idea is quite standard in the turbulence studies and was exploited in particular by Kazantsev \cite{Kazantsev1968}. Another point is that the method does not require a 
particular distribution for matrices $\hat A_i$ (e.g. an isotropic one). Thus
it can be applied in the anisotropic case as well.

The most challenging step in the above algorithm is
averaging of the matrix exponential. It can be
computed explicitly in rare cases only. In practical
cases, one has to consider $\tau$ as a small
parameter and approximate the matrix exponential
with the Taylor series. Below we elaborate it in a greater details.

Suppose the matrix $\hat A$ has a multivariate Gaussian distribution with zero mean and some covariance matrix $B$ of size $k^2\times k^2$ defined as follows:
\begin{equation}
B_{k(i-1)+j, k(l-1)+m} = \langle \hat A_{ij} \hat A_{lm} \rangle \, .
\label{corr_matrix}
\end{equation}
This assumption has two reasons. First, it is common
to define statistical properties of a random flow
via the correlation tensor. Thus it is convenient to
suppose the Gaussian distribution which is
completely defined by this tensor. Second,
for numerical experiments it is also convenient to
simulate Gaussian distribution given the correlation
tensor.

Once the distribution of matrix $\hat A$ is specified, it is also
specified for the distribution of the extended matrix $\doublehat A$ according to Eq.~(\ref{A_hat}). Now one can approximate the mean matrix exponential as follows:
\begin{equation}
\langle \exp(\doublehat A\tau) \rangle = I + \langle {\doublehat A} \rangle \tau + \frac{1}{2} \langle \doublehat A^2 \rangle \tau^2 + ...
\end{equation}

In the Gaussian case, calculation of $\langle \doublehat A^n \rangle$ for any $n$ can be performed as follows. The idea is to represent components of the matrix $\hat A$ (and, hence, of the matrix $\doublehat A$) as
a linear combination of independent standard random normal variables. The benefit is that averaging  $\doublehat A^n$ we only need to count degrees and
coefficients for these variables. This is a routine
algorithmic procedure that can be implemented in a computer program.

Here we recall a standard way to represent a
multivariate Gaussian vector in terms of independent
standard random normal variables. Given a covariance matrix $B$ (with a full rank for simplicity) we decompose it into a
product $B = L L^T$ (e.g. using the 
the Cholesky decomposition \citep{GoluVanl96}).
Now let $\eta$ be a vector of independent
standard random normal variables (i.e. with
a unit covariance matrix), then the vector $L\eta$
has the covariance matrix $\hat B$.

In the next sections
we consider application of the method to 2D and 3D
induction equations and obtain an approximation for
growth rate $\gamma_2$ assuming Gaussian distribution for derivatives of the velocity field.

\section{Basic equations}

We start out from the induction equation that reads
\begin{equation}
    {{\partial {\bf H} \over {\partial t}} + ({\bf v} \nabla} {\bf H}) = ({\bf H } \nabla) {\bf v} - \eta \, {\rm curl} \, {\rm curl}\, {\bf H} \, ,
    \label{ind}
\end{equation}
where $\bf H$ is magnetic field strength, $\bf v$ is velocity, $\eta$ is magnetic diffusivity considered as a space independent scalar quantity. According to our aims we neglect the the Ohmic losses  and use Lagrangian approach to obtain from Eq.~(\ref{ind})
\begin{equation}
    {{d \bf H} \over {dt}} = A \bf H\, ,
\label{largange}
\end{equation}
where $A$ is a matrix of derivatives of the velocity field, i.e. $ A_{ij} = \partial v_i/ \partial x_j$ and
derivatives are taken at appropriate points of a given Lagrangian trajectory. 

We consider $\hat A_{ij}$ as representative of random processes with the renovations at time instants $\tau$, $2 \tau$, $3\tau$ and so on. Generalization based on more
realistic models of the memory decay is also possible. In
particular, one can admit that $\hat A_{ij}$ between the
renovation times are random processes with specific memory
times $\tau^*$ independent on $\tau$,  as well as
suggest that the renovation instants are random and can be
considered as a Poisson process \cite{2000LS}.

Spatial structure of the random velocity field $\bf v$ will be defined via the two-point correlation tensor
$R_{ij}({\textbf{x}-\textbf{y}}) = \langle v_i({\rm{\textbf{x}}})v_j({\rm{\textbf{y}}}) \rangle$. We also assume $\langle {\bf v} \rangle=0$, however, non-vanishing mean velocity can be included in consideration as well following \cite{Tomin}. Further details depend on the particular form of the correlation tensor and are specific for 3D and 2D cases.

\section{There-dimensional isotropic model}

For statistically homogeneous and isotropic
velocity field $R_{ij}$ reads  (e.g. \cite{1953B}, \cite{monin_statistical_2007})
\begin{equation}
R_{ij}({\rm{\textbf{r}}}) = f(r)\delta_{ij} + \frac{r}{2}f'(r)\left ( 
\delta_{ij} - \frac{r_ir_j}{r^2} 
\right ) +
\gamma(r)\varepsilon_{ijk}r_k\,,
\label{corr_comp}
\end{equation}
where $\textbf{r} = \textbf{x} - \textbf{y}$, $r=\|\textbf{r}\|$ and
$f$ is the longitudinal correlation function.
It is common to consider the mirror-symmetric model,
which implies $\gamma(r)=0$.
Here we do not rely on this assumption but
show that the term $\gamma(r)$ does not affect the
correlation tensor of derivatives of the velocity field $\bf v$.

Following \cite{SI15} we normalize (for small $r$) the correlation function
in a way that
$\langle v_i({\rm{\textbf{x}}}) v_i({\rm{\textbf{y}}}) \rangle \sim v^2(1-(r/l)^2)$, where
$l$ defines the scale of random vortexes, $v$ is
the the rms velocity of the flow. To be specific, we use below 
$f(r) = (v^2/3)\exp(-3r^2/5l^2)$. In other words, we do not include the turbulent cascade in our model. 

To obtain the correlation tensor of velocity  derivatives $\partial v_i /\partial x_j$ we compute
second order 
partial derivatives
$\partial^2/\partial x_m\partial y_n$ of
velocity field correlations
$\langle v_i({\rm{\textbf{x}}})v_j({\rm{\textbf{y}}}) \rangle = R_{ij}(\rm{\textbf{r}})$ and let $\rm{\textbf{r}}\to 0$. Using the notation introduced in Eq.~(\ref{corr_matrix}) we obtain after neglecting the scale factor $v^2/5l^2$, the following
the correlation matrix $ B$:
\begin{equation}
 B = \begin{pmatrix}
2 & 0 &  0 & 0 & -1 &  0 &  0 &  0 & -1 \\
0 &  4 &  0 & -1 &  0 &  0 &  0 &  0 &  0 \\
0 &  0 &  4 &  0 &  0 &  0 & -1 &  0 &  0 \\
0 & -1 &  0 &  4 &  0 &  0 &  0 &  0 &  0 \\
-1 &  0 &  0 &  0 &  2 &  0 &  0 &  0 & -1 \\
0 &  0 &  0 &  0 &  0 &  4 &  0 & -1 &  0 \\
0 &  0 & -1 &  0 &  0 &  0 &  4 &  0 &  0 \\
0 &  0 &  0 &  0 &  0 & -1 &  0 &  4 &  0 \\
-1 &  0 &  0 &  0 & -1 &  0 &  0 &  0 &  2
\end{pmatrix} \, .
\label{corr_B}
\end{equation}

Computing $B$ we note that $\gamma(r)$ disappears in the correlation
tensor of velocity  derivatives. Indeed, taking partial derivatives we obtain 
\begin{equation}
\begin{split}
    \frac{\partial^2}{\partial x_m \partial y_n}\varepsilon_{ijk}r_k\gamma =
    \varepsilon_{ijk}\frac{\partial}{\partial x_m} \left(-\delta_{kn}\gamma + r_k\frac{\partial}{\partial y_n}\gamma\right) = \\
    =\varepsilon_{ijk}\left(-\delta_{kn}\frac{\partial}{\partial x_m}\gamma + \delta_{km}\frac{\partial}{\partial y_n}\gamma + r_k\frac{\partial^2}{\partial x_m \partial y_n}\gamma\right) \, .
\label{asym}
\end{split}
\end{equation}
Bearing in mind that  for any smooth function
$\gamma(r)$ first derivatives
$\partial/\partial x_m$ and
$\partial/\partial y_n$
vanish  at $r=0$, we see that $\gamma$ vanishes in the correlations under discussion.

In other words, the mirror asymmetry does not contribute to the evolution of the frozen-in magnetic field. This conclusion looks quite unexpected because effects of mirror asymmetry of the flow are known to play a crucial role at the subsequent stages of the dynamo action \cite{KR}. 

An explicit form of the Gaussian matrix $\hat A$ that corresponds to the correlation matrix $B$
is as follows:
\begin{equation}
\hat{A} = \frac{v}{\sqrt{5}l}\begin{pmatrix}
\sqrt{2}\eta_1 &
2\eta_2 & 
2\eta_3 \\

\frac{-\eta_2 + \sqrt{15}\eta_4}{2} &
\frac{-\eta_1 + \sqrt{3}\eta_5}{\sqrt{2}} & 
2\eta_6 \\

\frac{-\eta_3+ \sqrt{15}\eta_7}{2} &
\frac{-\eta_6+ \sqrt{15}\eta_8}{2} &
\frac{-\eta_1 - \sqrt{3}\eta_5}{\sqrt{2}}

\end{pmatrix} \,.
\label{A_eta}
\end{equation}
Here $\eta_1$, ..., $\eta_8$ are independent
standard random normal variables. Note that
only 8 variables are required to define the matrix $\hat A$. Due to the incompressibility condition i.e., ${\rm div} {\bf v} =0$,
the matrix $B$ has an incomplete rank of 8.

Using  Eq.~(\ref{A_hat}) and (\ref{A_eta}) we
completely specify the extended matrix $\doublehat A$.
To estimate the leading eigenvalue of the mean
 matrix exponential consider the following
characteristic equation 
\begin{equation}
\textrm{det}(\langle \exp(\doublehat{A}\tau) \rangle - \lambda I) = 0 \, .
\end{equation}
Approximating the matrix exponential with a Taylor series
for small $\tau$ and taking into account that 
$\langle \doublehat{A}^n \rangle=0$ for any odd $n$,
we obtain 
\begin{equation}
\textrm{det}\left(I(1-\lambda) + \frac{1}{2}\langle \doublehat{A}^2 \rangle\tau^2 +  \frac{1}{4!}\langle \doublehat{A}^4 \rangle\tau^4 + ...\right) = 0 \, .
\end{equation}
At this point we recall that the matrix $\doublehat{A}$ has the scale factor $v/l$. Thus the Taylor series can be naturally written in terms of the Strouhal number $s=\tau v/l$. Note that canonically one denotes the Strouhal number as $St$, in this work we use $s$ to simplify
analytical expressions. Note also that the Strouhal number is often defined as $St^{-1}$. With the consideration of these remarks and by solving the characteristic equation we obtain an $O(s^8)$ approximation for eigenvalues:
\begin{equation}
\begin{split}
    &\lambda_1 = 1 + 2s^2 + \frac{9}{10}s^4 + \frac{149}{500}s^6 + O(s^8) \, ,\\
    &\lambda_{2, ..., 7} = 1 - \frac{2}{5}s^2 + \frac{1}{2}s^4 - \frac{187}{7500}s^6 +  O(s^8) \, ,\\
    &\lambda_{8,\, 9} = 1 + \frac{1}{15}s^4 + \frac{8}{9000}s^6 + O(s^8) \, .
\end{split}
\label{roots}
\end{equation}
In order to obtain the growth rate $\gamma_2$, one
should take the logarithm of the leading eigenvalue $\lambda_1$
with a factor $1/4\tau$. This yields
\begin{equation}
    \gamma_2 = \frac{1}{2}s - \frac{11}{40}s^3 + \frac{1747}{6000}s^5 + O(s^7) \, 
\label{gamma_rate_3d}
\end{equation}
up to a scale factor of $v/l$. It is possible
to get several higher order approximations, however, we do not show them due
to large coefficients. Of course, we recover
the first-order estimation $\gamma_2=1/2[v/l]$ found in \cite{SI15}
(see Appendix A for more details).

In Fig.~\ref{rates_plot_3d} the above analytical result is compared 
with numerical estimation of the matrix exponential and corresponding value of $\gamma_2$ for various Strouhal numbers $s=\tau v/l$. Specifically, for each $s$ we
generate $10^5$ realizations of random matrices $\doublehat A$, evaluate matrix
exponential $\exp(\doublehat
A\tau)$ and obtain sample averaged matrix
exponential $\langle\exp(\doublehat
A\tau)\rangle$. Then we numerically derive the leading eigenvalue $\lambda$ of $\langle\exp(\doublehat
A\tau)\rangle$ and obtain
$\gamma_2=(1/4s)\lambda$.
Fig.~\ref{rates_plot_3d} shows that for 
small $s$ ($s<0.6$) analytical approximation (\ref{gamma_rate_3d}) and numerical estimation of $\gamma_2$
are in good agreement.

\begin{figure}[h]
\centering
\includegraphics[width=0.48\textwidth]{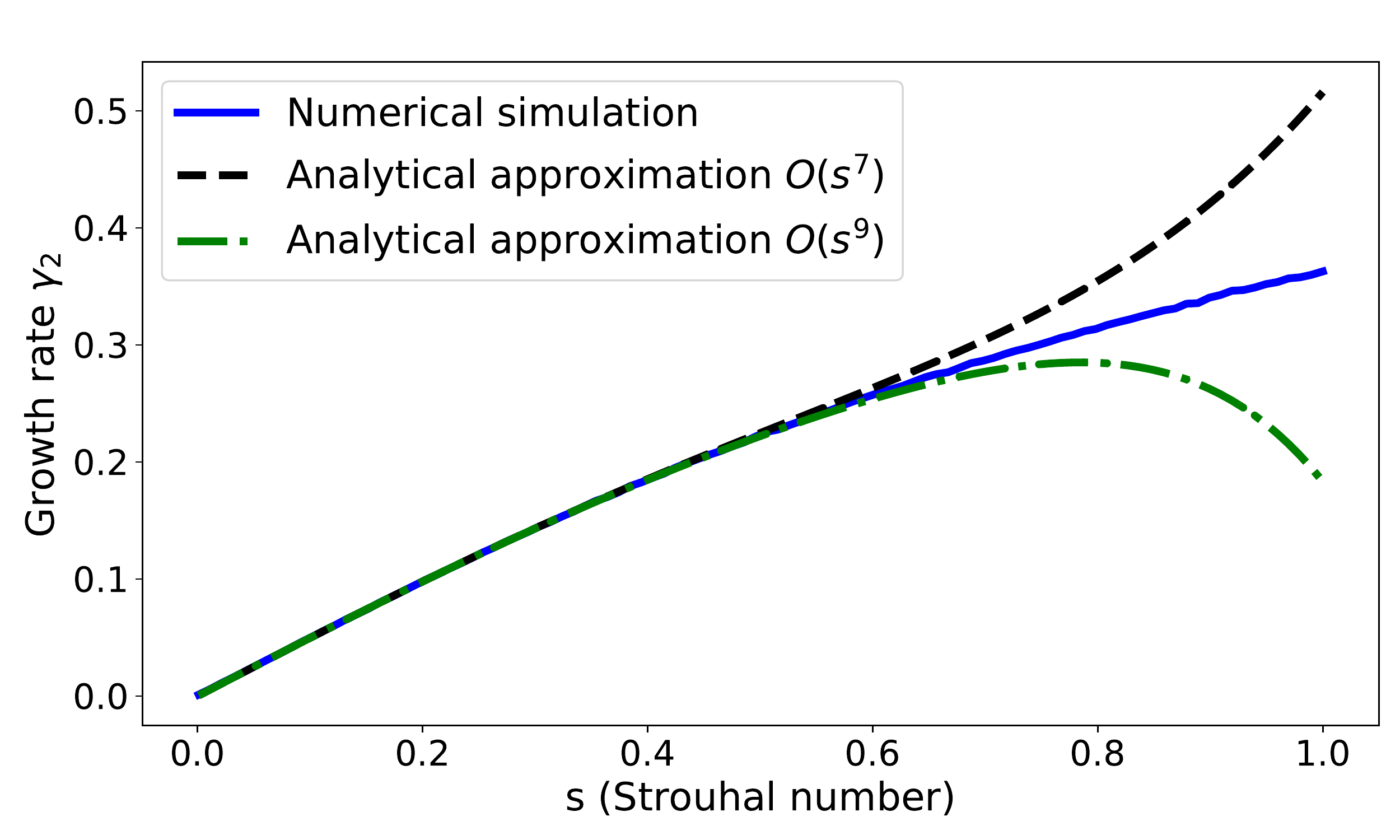}
\caption{Numerical and analytical approximation of $\gamma_2$ for small Strouhal numbers in the isotropic 3D flow model.}
\label{rates_plot_3d}
\end{figure}

One can also note in Fig.~\ref{rates_plot_3d}
that two successive approximations, $O(s^7)$ and $O(s^9)$,
give opposite behaviour for $s\to+\infty$. This is due to the fact that the coefficients in (\ref{gamma_rate_3d}) have
alternating signs.

\section{Two-dimensional isotropic model}

Here we consider the 2D case that allows explicit calculation of matrix exponents and a bit deeper analytical investigation. Correlation tensor for 2D isotropic flow reads
\begin{equation}
R_{ij}({\rm{\textbf{r}}}) = f(r)\delta_{ij} + rf'(r)\left ( 
\delta_{ij} - \frac{r_ir_j}{r^2} 
\right ) \, .
\label{corr_2В}
\end{equation}
Note that the mirror asymmetry effects can not be included in this model because the definition of antisymmetric tensor $\varepsilon_{ijk}$ presumes 3D.
Normalization condition
$\langle v_i({\rm{\textbf{x}}}) v_i({\rm{\textbf{y}}}) \rangle \sim v^2(1-(r/l)^2)$ implies
$f(r) = (v^2/2)\exp(-r^2/2l^2)$. Thus a correlation
matrix for $\partial v_i/ \partial x_j$ is
\begin{equation}
B = \frac{v^2}{2l^2}\begin{pmatrix}
1 & 0 &  0 & -1 \\
0 &  3 &  -1 & 0 \\
0 &  -1 &  3 &  0 \\
-1 & 0 &  0 &  1 
\end{pmatrix} \, .
\label{corr_B_2d}
\end{equation}
Corresponding
Gaussian matrix $\hat A$ is
\begin{equation}
\hat A = \frac{v}{l}\begin{pmatrix}
\eta_1/\sqrt{2} & \sqrt{3}\eta_2/\sqrt{2} \\
-\eta_2/\sqrt{6} + 2\eta_3/\sqrt{3} &  -\eta_1/\sqrt{2}
\end{pmatrix} \, 
\label{A_2d}
\end{equation}
where $\eta_1$, $\eta_2$, $\eta_3$ are independent standard random normal variables. Taking into account that $\hat A_{11} = - \hat A_{22}$,
the extended matrix $\doublehat{A}$ reads
\begin{equation}
\doublehat{A} = \begin{pmatrix}
2{\hat A}_{11} & {\hat A}_{12} & {\hat A}_{12} & 0 \\
{\hat A}_{21} & 0 & 0 & {\hat A}_{12} \\
{\hat A}_{21} & 0 & 0 & {\hat A}_{12} \\
0 & {\hat A}_{21} & {\hat A}_{21} & -2{\hat A}_{11}
\end{pmatrix} \,.
\label{A_2d_ext}
\end{equation}

Matrix exponential $\doublehat{E} = \exp\doublehat{A}\tau$
can be computed explicitly.
For simplicity, let $\tau=1$ and denote
$\varphi=\det \hat A = -\hat A_{11}^2 - \hat A_{12}\hat A_{21}$. A  straightforward calculation then yields
\begin{equation}
\begin{split}
&\doublehat{E}_{11} =  \cos^2\sqrt\varphi + \frac{\hat A_{11}^2}{\varphi}\sin^2\sqrt\varphi + \frac{{\hat A}_{11}}{\sqrt\varphi}\sin2\sqrt\varphi \,,\\
&\doublehat{E}_{12} = \doublehat{E}_{13} =  \frac{{\hat A}_{11}{\hat A}_{12}}{\varphi}\sin^2\sqrt\varphi + \frac{{\hat A}_{12}}{2\sqrt\varphi}\sin2\sqrt\varphi \,,\\
&\doublehat{E}_{14} =  \frac{{\hat A}_{12}^2}{\varphi}\sin^2\sqrt\varphi \,,\\
&\doublehat{E}_{21} = \doublehat{E}_{31} =  \frac{{\hat A}_{11}{\hat A}_{21}}{\varphi}\sin^2\sqrt\varphi + \frac{{\hat A}_{21}}{2\sqrt\varphi}\sin2\sqrt\varphi \,,\\
&\doublehat{E}_{23} = \doublehat{E}_{32} =  \frac{{\hat A}_{12}{\hat A}_{21}}{\varphi}\sin^2\sqrt\varphi \,,\\
&\doublehat{E}_{22} = \doublehat{E}_{33} = 1 + \doublehat{E}_{23} \,,\\
&\doublehat{E}_{24} = \doublehat{E}_{34} =  -\frac{{\hat A}_{11}{\hat A}_{12}}{\varphi}\sin^2\sqrt\varphi + \frac{{\hat A}_{12}}{2\sqrt\varphi}\sin2\sqrt\varphi \,,\\
&\doublehat{E}_{41} = 
\frac{{\hat A}_{21}^2}{\varphi}\sin^2\sqrt\varphi \,,\\
&\doublehat{E}_{42} = \doublehat{E}_{43} =  -\frac{{\hat A}_{11}{\hat A}_{21}}{\varphi}\sin^2\sqrt\varphi + \frac{{\hat A}_{21}}{2\sqrt\varphi}\sin2\sqrt\varphi \,,\\ 
&\doublehat{E}_{44} =  \cos^2\sqrt\varphi + \frac{\hat A_{11}^2}{\varphi}\sin^2\sqrt\varphi - \frac{{\hat A}_{11}}{\sqrt\varphi}\sin2\sqrt\varphi \,.
\label{expB2_ext}
\end{split}
\end{equation}

Averaging $\doublehat E$ and using symmetry of probability distributions we obtain
\begin{equation}
\begin{split}
&\langle \doublehat{E}_{12}\rangle =
\langle \doublehat{E}_{13}\rangle =
\langle \doublehat{E}_{21}\rangle =
\langle \doublehat{E}_{31}\rangle =\\
&=\langle \doublehat{E}_{24}\rangle =
\langle \doublehat{E}_{34}\rangle =
\langle \doublehat{E}_{42}\rangle =
\langle \doublehat{E}_{43}\rangle = 0 \, .
\end{split}
\label{zero_property}
\end{equation}
In matrix terms, this means that only the following matrix elements denoted by the star symbol can be nonzero:
\begin{equation}
\langle {\cal\hat E} \rangle = \begin{pmatrix}
* & 0 &  0 & * \\
0 &  * &  * & 0 \\
0 &  * &  * &  0 \\
* & 0 &  0 &  * 
\end{pmatrix} \, .
\end{equation}
In terms of the vector $\langle \bf Z \rangle$, this means that we can separate a closed equation for the 
vector $(\langle Z_{11} \rangle, \langle Z_{22}\rangle) = (\langle H_{1}^2\rangle, \langle H_{2}^2\rangle)$. It remains to note that due to the symmetry of probability
distributions $\langle ({\hat A}_{11}/\sqrt\varphi)\sin2\sqrt\varphi\rangle =0$ we obtain
$\langle \doublehat{E}_{11}\rangle =\langle \doublehat{E}_{44}\rangle $
as well as
$\langle \doublehat{E}_{14}\rangle =\langle \doublehat{E}_{41}\rangle $. 
Then, denoting
\begin{equation}
    {\lambda}=\langle \doublehat{E}_{11} + \doublehat{E}_{14}\rangle = \langle \cos^2\sqrt\varphi + \frac{\hat A_{11}^2 + \hat A_{12}^2}{\varphi}\sin^2\sqrt\varphi \rangle \,,
\label{lambda_exp}
\end{equation}
we arrive at a closed equation for magnetic energy
\begin{equation}
\langle H_1^2 + H_2^2\rangle((n+1)\tau)  
= 
{\lambda}
\langle H_1^2 + H_2^2\rangle(n\tau) \, .
\label{H2equation}
\end{equation}
An approximation of (\ref{lambda_exp}) for small $s$ reads
\begin{equation}
   \lambda = 1 + 2s^2+
   \frac{2}{3}s^4 + 
   \frac{4}{9}s^6 + O(s^8) \,.
\label{lambda_iso_2d}
\end{equation}
Corresponding growth rate is (omitting scale factor $v/l$)
\begin{equation}
\gamma_2 = \frac{1}{4\tau}\ln\lambda = \frac{1}{2}s -\frac{1}{3}s^3 +\frac{4}{9}s^5 +  O(s^7) \,.
\label{gamma2_2d}
\end{equation}

\begin{figure}[h]
\centering
\includegraphics[width=0.45\textwidth]{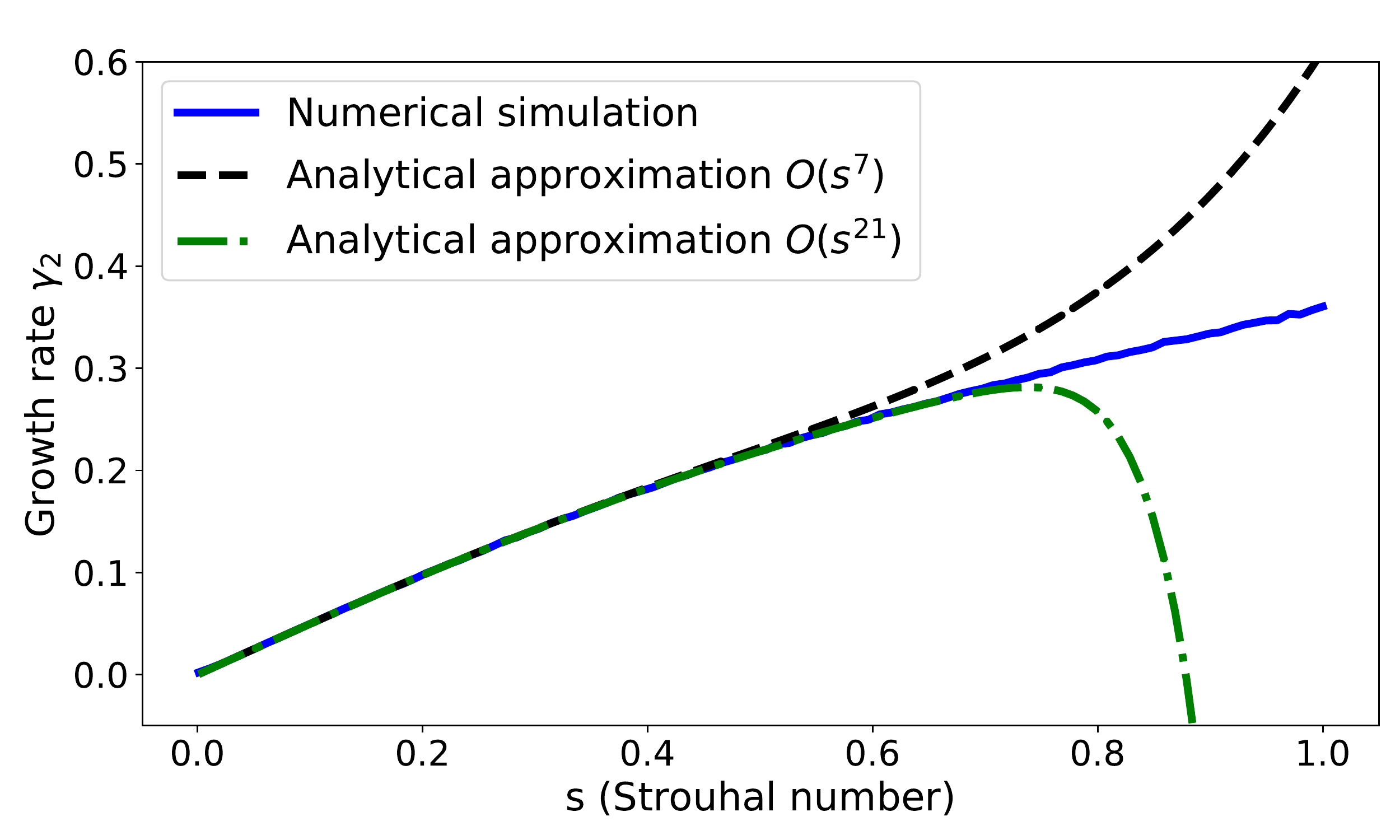}
\caption{Numerical and analytical approximation of
$\gamma_2$ for small Strouhal numbers in the isotropic 2D flow model.}
\label{rates_plot_2d}
\end{figure}

In Fig.~\ref{rates_plot_2d} we compare the analytical and numerical
approximation of $\gamma_2$ for small Strouhal numbers $s$ following
the scheme introduced in the 3D case.

We observe in Fig.~\ref{rates_plot_2d} that
the 2D model reproduces the features of the 3D model
noted in Fig.~\ref{rates_plot_3d}. Also in agreement
with 3D case, the coefficients
in analytical approximation (\ref{gamma2_2d}) have alternating signs.

\section{Anisotropic flow}

\subsection{Three-dimensional axisymmetric model}

In order to introduce some anisotropy in the model we consider axisymmetric turbulence model developed by Chandrasekhar \cite{Chandrasekhar}.
In particular, we consider the first axis as an axis of symmetry (or a preferred direction) and a correlation tensor of
the form
\begin{equation}
    R_{ij} = \varepsilon_{jkm} \frac{\partial q_{im}}{\partial r_k} \, ,
\label{Rij_axi}
\end{equation}
where $q_{im}$ is a skew axisymmetric tensor defined as
\begin{equation}
    q_{im} = r_k(\varepsilon_{imk}Q_1 +
    \varepsilon_{i1k}(\delta_{1m}Q_2 + r_m Q_3)) \, .
    \label{q_im}
\end{equation}
Due to the symmetry and zero divergence of the correlation
tensor, the functions $Q_i$ are not independent. In particular, 
\begin{equation}
    Q_3 = \left( \frac{\partial}{\partial r_1} -
    \frac{r_1}{r_3}\frac{\partial}{\partial r_3} \right)Q_1 \, .
\end{equation}
This implies that only two functions, $Q_1$ and $Q_2$, should
be specified.

Of course, the isotropic case considered earlier
in this paper should be a special case of axisymmetric one.
To be specific, we follow the model proposed by Khavaran \cite{Khavaran} 
that allows such a generalization. In particular, we assume
\begin{equation}
\begin{split}
    &Q_1 = -\frac{1}{2}v_1^2\exp\left[-\frac{3}{5}\left(\frac{r_1^2}{l_1^2}+\frac{r_2^2+r_3^2}{l_2^2}\right)\right] \, ,\\
    &Q_2 = -(v_2^2 - v_1^2)\exp\left[-\frac{3}{5}\left(\frac{r_1^2}{l_1^2}+\frac{r_2^2+r_3^2}{l_2^2}\right)\right] \, .
\end{split}
\label{Q12}
\end{equation}
Here $v_1$ and $l_1$ are mean velocity and length scale along
the first direction (which is the axis of symmetry), while $v_2$ and $l_2$
are corresponding values in the orthogonal plane.
Let's define $u=v_2^2/v_1^2$ and 
$L=l_2^2/l_1^2$ that can be considered as 
the power
and wavevector anisotropy. Then
\begin{equation}
\begin{split}
    &Q_1 = -\frac{1}{2}v_1^2\exp\left[-\frac{3}{5l_1^2}\left(r_1^2+\frac{r_2^2+r_3^2}{L}\right)\right] \, ,\\
    &Q_2 = (1-u) v_1^2 \exp\left[-\frac{3}{5l_1^2}\left(r_1^2+\frac{r_2^2+r_3^2}{L}\right)\right] \, .
\end{split}
\label{Q12k}
\end{equation}
Note that $u=L=1$ results in the isotropic model
(\ref{corr_2В}).

A correlation matrix for $\partial v_i / \partial x_j$
has the same structure (i.e. positions of nonzero elements) as matrix $B$ in the isotropic model (see Eq.~(\ref{corr_B})). Omitting a common multiplier 
$3v_1^2/5l_1^2$, the nonzero elements are:
\begin{equation}
\begin{split}
&B_{11} = 2 \, ,\\
&B_{22} = B_{33} = \frac{4}{L} \, ,\\
&B_{44} =  B_{77} = 2u+4- \frac{2}{L} \, ,\\
& B_{55} =  B_{99} = \frac{2u}{L} \, ,\\
& B_{66} =  B_{88} = \frac{2(3u-L)}{L} \, , \\
& B_{42} =  B_{51} =  B_{73} =  B_{91} = -1 \, ,\\
& B_{24} =  B_{15} =  B_{37} =  B_{19} = -1  \, ,\\
& B_{68} =  B_{86} =  B_{59} =  B_{95} = \frac{L-2u}{L} \, .
\end{split}
\end{equation}
Note again that $u=L=1$ results in the correlation
matrix (\ref{corr_B}). Also $v^2=3v_1^2$ and $l^2=l_1^2$ in the isotropic model.

A straightforward calculation shows that up to a scale factor $\sqrt{3}v_1/\sqrt{5}l_1$, components of
the corresponding Gaussian matrix $\hat A$ are
\begin{equation}
\begin{split}
&\hat{A}_{11}  = \sqrt{2}\eta_1 \, ,\\
&\hat{A}_{12} = \frac{2}{\sqrt{L}}\eta_2 \, , \\
&\hat{A}_{13} = \frac{2}{\sqrt{L}}\eta_3  \, ,\\
&\hat{A}_{21} = \frac{-\sqrt{L}}{2}\eta_2  
+\sqrt{2u + 4 - \frac{2}{L} - \frac{L}{4}}\eta_4  \, ,\\
&\hat{A}_{22} = -\frac{1}{\sqrt{2}}\eta_1 + \sqrt{\frac{4u-L}{2L}}\eta_5 \, , \\
&\hat{A}_{23} = \sqrt{\frac{2(3u-L)}{L}}\eta_6  \, ,\\
&\hat{A}_{31} = \frac{-\sqrt{L}}{2}\eta_3  
+\sqrt{2u + 4 - \frac{2}{L} - \frac{L}{4}}\eta_7  \, ,\\
&\hat{A}_{32} =  \frac{L-2u}{\sqrt{2L(3u-L)}}\eta_6 +\sqrt{\frac{(8u-3L)(4u-L)}{2L(3u-L)}}\eta_8 \, ,\\
&\hat{A}_{33} = -\frac{1}{\sqrt{2}}\eta_1 - \sqrt{\frac{4u-L}{2L}}\eta_5\, .
\end{split}
\label{A_matrix}
\end{equation}

Note that it is not possible to have any arbitrary positive $L$ and $u$ as the arguments under the radical signs should be real and positive. Considering the expressions under the square roots in (\ref{A_matrix}), we conclude that
\begin{equation}
\begin{split}
&2u + 4 - \frac{2}{L} - \frac{L}{4} \geqslant 0 \, ,\\
& 4u-L  \geqslant 0 \, ,\\
& 3u-L  \geqslant 0 \, ,\\
& 8u-3L  \geqslant 0 \, .
\end{split}
\label{inequalities}
\end{equation}

The next step includes consideration of extended 
system and computation of eigenvalues. 
After some algebra 
we obtain $O(s^8, u^2, L^2)$ approximation of the leading eigenvalue:
\begin{equation}
\begin{split}
    &\lambda_1 = 1 + 2\left(1 + \frac{2\varepsilon}{3}\right)s^2 + \\ &+\frac{9}{10}\left(1+\frac{4\varepsilon}{3}\right)s^4
    + \frac{149}{500}\left(1+2\varepsilon\right)s^6 \, ,
\end{split}
\label{axi_leading}
\end{equation}
where $\varepsilon=u - L$, $s=\sqrt{3}v_1\tau/l_1$. Note that  the Strouhal number $s$ depends on $v_1$ and quantities in (\ref{Q12}) and (\ref{Q12k}) are also
normalized to $v_1$. Thus the contribution of the anisotropy parameters $u$ and $L$ and the rms velocity to the growth rate are separated.
Note also that for $u=L$ we obtain the same
result as for  $u=L=1$ (isotropic model).
This looks reasonable because similar change of
velocity and length does not change the Strouhal number, which is
responsible for the growth rate.

Rewriting the eigenvalue (\ref{axi_leading}) in terms
of growth rate we obtain:
\begin{equation}
\begin{split}
    &\gamma_2 = \frac{1}{2}\left(1+\frac{2\varepsilon}{3}\right)s + \\
    &-\frac{11}{40}\left(1+\frac{4\varepsilon}{3}\right)s^3 + \frac{1747}{6000}\left(1+2\varepsilon\right)s^5 \, .
\end{split}
\label{}
\end{equation}

\begin{figure}[h]
\centering
\includegraphics[width=0.48\textwidth]{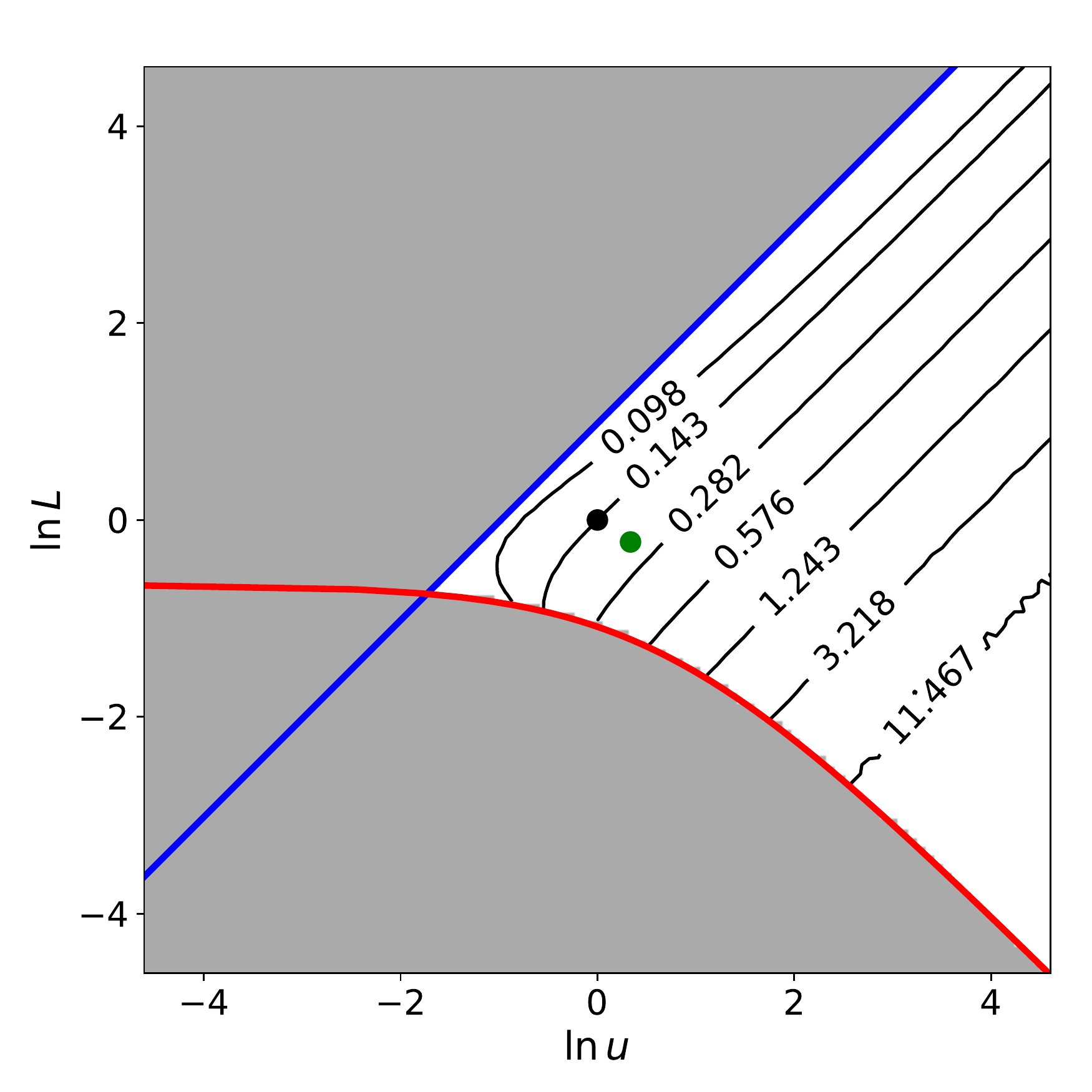}
\caption{Contour plot of the growth rate $\gamma_2$ distribution in the axisymmetric 3D flow model
for fixed value $s=0.3$ and various
anisotropy parameters $u$, $L$. Note the log scale of both axes. Black dot corresponds to the isotropic case $u=L=1$,
green dot corresponds to the case $u=1.4$, $L=0.8$ shown in Fig.~\ref{axi_rates_3d_plot}. Gray area shows forbidden
$u$ and $L$ values for 3D model. Red line corresponds
to the first inequality in (\ref{inequalities}), blue line corresponds to the last inequality in (\ref{inequalities}).}
\label{axi_plane_3d_plot}
\end{figure}

To demonstrate (Fig.~\ref{axi_plane_3d_plot}) the role of the anisotropy parameters we simulate $10^5$
realizations of random matrix $\hat A$ for various $u$ and $L$ following
(\ref{A_matrix}), compute numerically the
leading eigenvalue of the sample-averaged  extended matrix $\langle \exp {\doublehat A}\tau \rangle$
and derive the corresponding growth rate $\gamma_2$. In these simulations the Strouhal
number $s$ remains fixed. We note in Fig.~\ref{axi_plane_3d_plot} that the growth
rate depends on the ratio $u/L$ rather than on particular values
of $u$ and $L$.

\begin{figure}[h]
\centering
\includegraphics[width=0.48\textwidth]{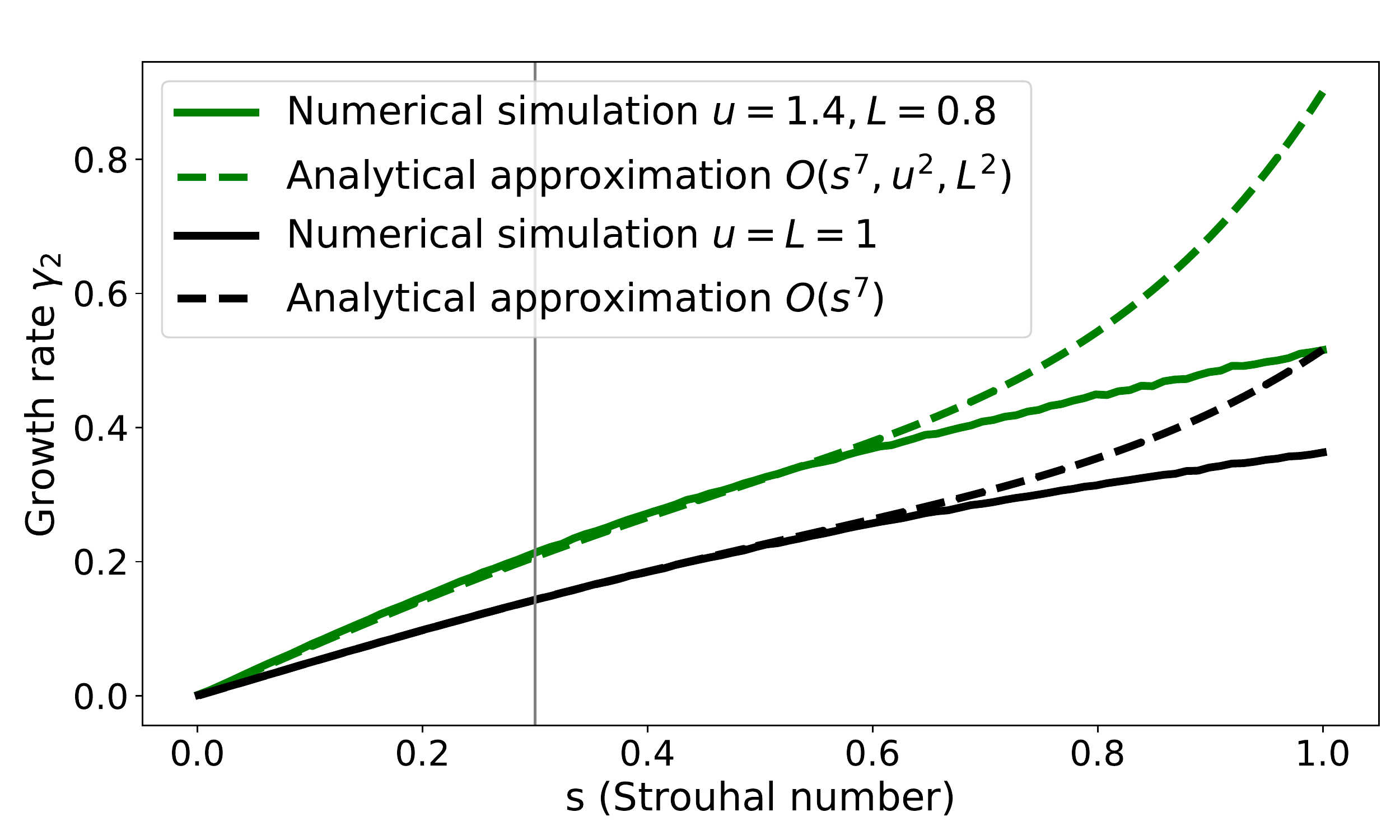}
\caption{Growth rate for the isotropic 3D flow model $u=L=1$ (black line)  in comparison to the anisotropy case $u=1.4$, $L=0.8$ (green line) for various Strouhal numbers $s$.
Dashed lines show corresponding analytical approximations. Vertical line shows a case $s=0.3$ given in Fig.~\ref{axi_plane_3d_plot}.}
\label{axi_rates_3d_plot}
\end{figure}

In contrast, in Fig.~\ref{axi_rates_3d_plot} we show 
results of similar numerical simulation against analytical approximation of the growth rate
for various Strouhal numbers $s$ and fixed anisotropy parameters
$u$ and $L$.
In particular, we compare $u=L=1$ (isotropic model) against the 
anisotripic model with slightly perturbed values $u$ and $L$. Similar to the isotropic case, we 
conclude that the analytical approximation
is accurate for small $s$ ($s<0.6$) only.

\subsection{Two-dimensional axisymmetric model}

Now we consider a 2D space. Our goal is to define a correlation tensor that could be compatible, at least conceptually, with the 3D axisymmetric model of Chandrasekhar \cite{Chandrasekhar}. 

We start with $R_{11}$ component. In 3D axisymmetric case
following (\ref{Rij_axi}) and (\ref{q_im}) we have
\begin{equation}
   R_{11} = -\frac{\partial (r_2Q_1)}{\partial r_2} - \frac{\partial (r_3Q_1)}{\partial r_3} \, .
\label{}
\end{equation}
It is natural to let
\begin{equation}
   R_{11} = -\frac{\partial (r_2Q_1)}{\partial r_2} \, ,
\label{R11}
\end{equation}
in 2D case, where $Q_1$, as in the case of (\ref{Q12}), is given by
\begin{equation}
   Q_{1} = -v_1^2\exp\left[-\frac{1}{2}\left(\frac{r_1^2}{l_1^2}+\frac{r_2^2}{l_2^2}\right)\right] \, .
\label{Q1}
\end{equation}

Now the incompressibility condition reads
\begin{equation}
   \frac{\partial R_{11}}{\partial r_1} +
   \frac{\partial R_{21}}{\partial r_2} =0 \, .
\label{}
\end{equation}
This implies that
\begin{equation}
   R_{21} = -\int \frac{\partial R_{11}}{\partial r_1} d r_2\, .
\label{}
\end{equation}
Taking into account (\ref{R11}) we obtain:
\begin{equation}
   R_{21} = \frac{\partial (r_2Q_1)}{\partial r_1} + C(r_1)\, .
\label{R21}
\end{equation}
Due to the symmetry of correlation tensor, $R_{21} = R_{12}$.

Note that in 3D case, according to (\ref{Rij_axi}) and (\ref{q_im}), we have
\begin{equation}
   R_{21} = R_{12} = -\frac{\partial (r_2Q_1)}{\partial r_1} \, .
\label{}
\end{equation}
Thus we assume $C(r_1)=0$ in 2D case.

In order to reconstruct the last component, $R_{22}$,
we consider again the incompressibility condition:
\begin{equation}
   \frac{\partial R_{12}}{\partial r_1} +
   \frac{\partial R_{22}}{\partial r_2} =0 \, .
\label{}
\end{equation}
It implies that
\begin{equation}
   R_{22} = -\int \frac{\partial R_{12}}{\partial r_1} d r_2\, .
\label{}
\end{equation}
Using (\ref{R21}) and (\ref{Q1}) we obtain:
\begin{equation}
   R_{22} = \frac{l_2^2}{l_1^2}\left(\frac{r_1^2}{l_1^2} - 1\right)Q_1 + \tilde{C}(r_1)\, .
\label{}
\end{equation}
To specify the function $\tilde{C}(r_1)$ we note that
$R_{22}(0) = v_2^2$, while $Q_1(0) = -v_1^2$.
Thus we assume
\begin{equation}
   \tilde{C}(r_1) = \left(v_2^2 - v_1^2\frac{l_2^2}{l_1^2}\right)
   \exp\left(-\frac{r_1^2}{2l_1^2}\right)\, .
\label{}
\end{equation}

The anisotropy parameters are again $v_2^2/v_1^2 = u$
and $l_2^2/l_1^2 = L$. A correlation matrix for 
$\partial v_i / \partial x_j$ is 
\begin{equation}
 B = \frac{v_1^2}{l_1^2}\begin{pmatrix}
1 & 0 &  0 & -1 \\
0 &  \frac{3}{L} &  -1 & 0 \\
0 &  -1 &  u + 2L &  0 \\
-1 & 0 &  0 &  1 
\end{pmatrix} \, .
\label{}
\end{equation}
In particular, isotropic case corresponds to $u=L=1$, $v^2=2v_1^2$,
$l^2=l_1^2$ and the matrix $ B$ takes the form (\ref{corr_B_2d}).

Gaussian matrix $\hat A$ that
corresponds to the correlation matrix $ B$ consists of
the following elements with common multiplier $v_1/l_1$:
\begin{equation}
\begin{split}
&\hat{A}_{11} = -\hat{A}_{22} = \eta_1 \,, \\
&\hat{A}_{12} = \sqrt{\frac{3}{L}}\eta_2 \,,\\
&\hat{A}_{21} = -\sqrt{\frac{L}{3}}\eta_2 +
\sqrt{\frac{3u+5L}{3}}\eta_3 \,.
\end{split}
\label{}
\end{equation}
Note that in contrast to 3D case, all positive $u$ and $L$ are possible.

Note that distributions of
$\hat{A}_{12}$ and $\hat{A}_{21}$
become different in contrast to 
the isotropic case. It implies  $\langle \doublehat{E}_{14}\rangle \ne \langle \doublehat{E}_{41}\rangle$ and instead of (\ref{H2equation})
we obtain
\begin{equation}
   \lambda_{1,2} = 
   \langle \doublehat{E}_{11}\rangle
   \pm \sqrt{\langle \doublehat{E}_{14}\rangle \langle \doublehat{E}_{41}\rangle}\, .
\label{lambda12_axi}
\end{equation}

Let $u\sim 1$, $L\sim 1$. We then obtain $O(s^8, u^2, L^2)$ approximation of the leading eigenvalue
\begin{equation}
   \lambda_1 = 1 + 2\left(1 + \frac{\varepsilon}{8}\right)s^2+
   \frac{2}{3}\left(1+\frac{\varepsilon}{4}\right)s^4 + 
   \frac{4}{9}\left(1 + \frac{3\varepsilon}{8}\right)s^6 \, ,
\label{lambda_axi}
\end{equation}
where $s=\sqrt{2}v_1\tau/l_1$, $\varepsilon=u-L$. Note again that
the Strouhal number $s$ is normalized to $v_1$ as well as $Q_1$ that allows a separation of the anisotropy and
the rms value contributions.

Correspondingly, an $O(s^7, u^2, L^2)$ approximation of the growth rate is (omitting a scale factor of $\sqrt{2}v_1/l_1$)
\begin{equation}
\gamma_2 = \frac{1}{2}\left(1 + \frac{\varepsilon}{8}\right)s -\frac{1}{3}\left(1+\frac{\varepsilon}{4}\right)s^3 +\frac{4}{9}\left(1 + \frac{3\varepsilon}{8}\right)s^5 \,.
\label{gamma2_2d_axi}
\end{equation}
Note that to a first approximation the anisotropy
parameters $u$ and $L$ contribute to the expression obtained only via the
difference $\varepsilon=u-L$
as we previously observed in 3D model.
Of course, (\ref{lambda_axi}) and then (\ref{gamma2_2d_axi}) can be obtained as well from  Taylor approximation of the matrix exponential $\doublehat{E} = \exp{\doublehat{A}\tau}$.

\begin{figure}[h]
\centering
\includegraphics[width=0.48\textwidth]{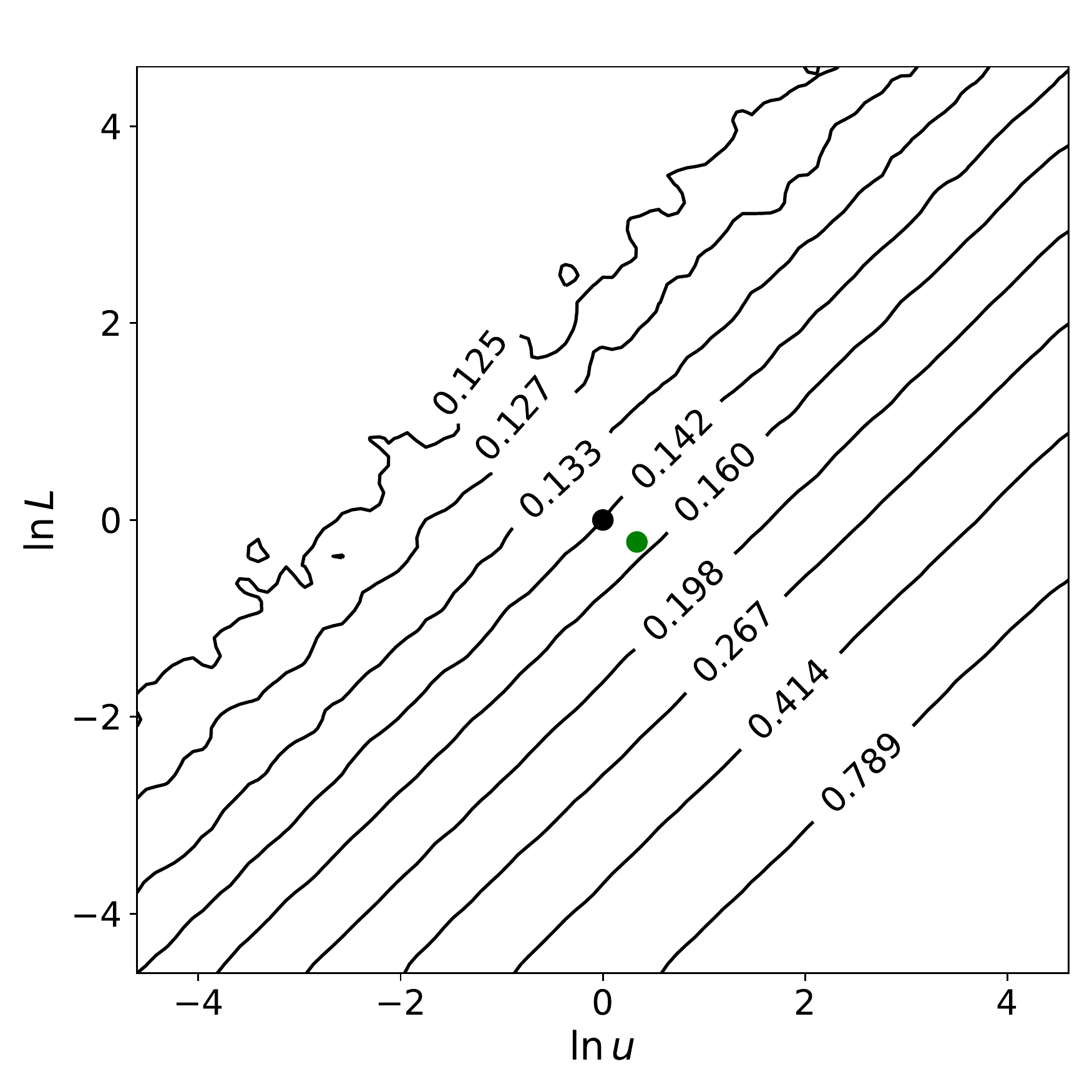}
\caption{Contour plot of the growth rate $\gamma_2$ distribution in the axisymmetric 2D flow model
for fixed value $s=0.3$ and various
anisotropy parameters $u$, $L$. Note the log scale of both axes. Black dot corresponds to the isotropic case $u=L=1$,
green dot corresponds to the case $u=1.4$, $L=0.8$ shown in Fig.~\ref{axi_rates_2d_plot}.}
\label{axi_plane_2d_plot}
\end{figure}

Fig.~\ref{axi_plane_2d_plot} shows 
a numerical estimation of the growth
rate according to (\ref{lambda12_axi}).
We find that plotting in log-log coordinates, the growth rate depends
only on the difference of $\ln u - \ln L$, or, equivalently, 
on the ratio of anisotropy parameters
$u/L$. For 
$u\sim 1$, $L\sim 1$ the ratio is 
approximated by $\varepsilon=u-L$ that is
in agreement with (\ref{gamma2_2d_axi}).

\begin{figure}[h]
\centering
\includegraphics[width=0.48\textwidth]{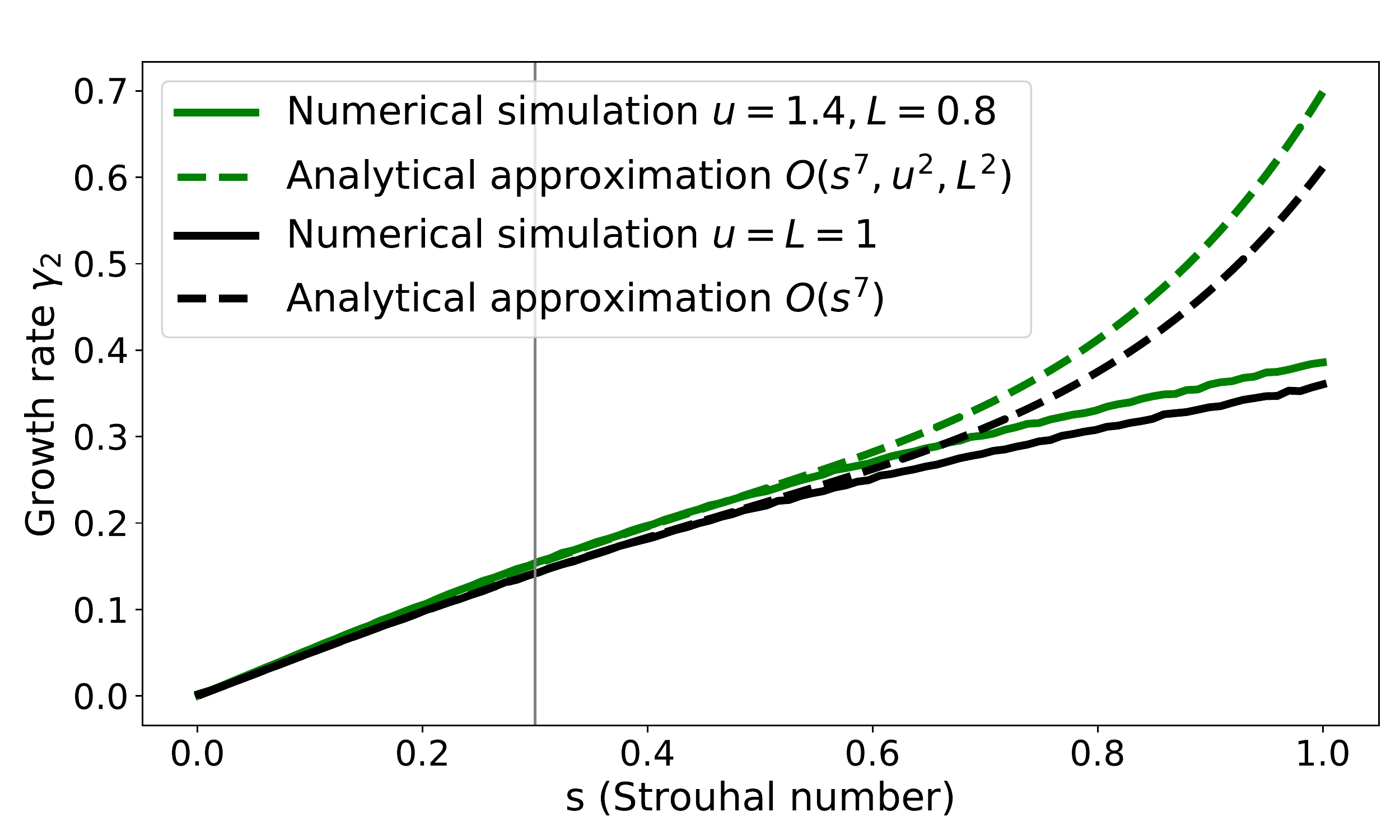}
\caption{Growth rate for the isotropic 2D flow model $u=L=1$ (black line) in comparison to the asymmetric case $u=1.4$, $L=0.8$ (green line) for various Strouhal numbers $s$.
Dashed lines show corresponding analytical approximations. Vertical line shows a case $s=0.3$ given in Fig.~\ref{axi_plane_2d_plot}.}
\label{axi_rates_2d_plot}
\end{figure}

\section{Conclusions and discussion}

We presented the generic approach to calculation of growth rate of higher statistical moments of a vector field
in the framework of the renovation model
of random flow. Then we applied this approach to investigate the magnetic energy  growth rate in various models where magnetic diffusion is neglected (in particular, at the initial stage of dynamo action).

Let us summarize the quantitative results and discuss its limits of applicability. We demonstrated that 
analytical approximations of the growth rate $\gamma_2$ are reasonable for quite
moderate Strouhal numbers only (up to $s=0.6$). 
For larger Strouhal numbers the analytical results
deviate substantially from numerical ones and become inapplicable.  However, practically relevant range of Strouhal numbers is rather uncertain
and remains a matter of discussion. For example,
some numerical experiments
give the Strouhal numbers of the order of 0.1 to 0.4 for
stellar convection zones (see \cite{Kapyla2006}).  At the same time, rough observational estimations from the solar granulation give $s\approx1$ \cite{Stix2002}.   Even higher estimates were derived from radio
observations of particular star
formation regions ($s^{-1}\approx 0.3$, \cite{Sobolev2018}) and observations of plasma flows in the solar corona ($s^{-1}\approx 0.2$, \cite{Samanta2019}). Note the inverse definition of the Strouhal number used is some papers.

Similar conclusions are obtained for the anisotropy effects, investigated in the framework of the Chandrasekhar's model of random flow with axisymmetric correlations. As expected, the energy growth rate turns out to depend on the details of anisotropic distribution. However, the dependence obtained is not very dramatic and extrapolation of isotropic results also gives at least qualitative understanding of the situation. This looks instructive for astrophysical applications. For example, interstellar turbulence in discs of spiral galaxies might be substantially anisotropic because the spatial scale of random motions is only few times smaller than the galactic disc thickness \cite{Betal96}. On the other hand, it would be too ambitious to insist that we know fine details of this anisotropy to use them in pragmatic dynamo models  in particular galaxies. Our result looks as a justification of conventional usage of isotropic models in such cases. 

We obtained two more or less unexpected results that can be summarized as follows. 

Firstly, the growth rates obtained do not depend on the mirror asymmetry of correlation tensor of turbulent velocities, while the basic idea of most important dynamo models (solar, stellar, galactic dynamos) are based on mirror asymmetry (e.g. \cite{Z83}). It means that the effects of mirror asymmetry become  important on later stages of dynamo action only. {\it Post factum}, this sounds reasonable.  Indeed, mirror asymmetry becomes important to avoid magnetic field cancellation of two oppositely directed  parts of  magnetic loop stretched by the flow. It requires a weak however finite Ohmic losses and can not be included in our consideration. Similar effects are known for the illustrative example of dynamo action in a stochastic flow in a Riemannian space \cite{Arnold1981}. 

Further the effects of mirror asymmetry are crucial for large-scale magnetic field excitation, while we investigate magnetic energy which can be presented as a small-scale magnetic field only. Effects of mirror asymmetry also present in small-scale dynamo but are much less crucial in comparison to large-scale ones \cite{FYu}. In any case, the result obtained implies that the statement that dynamo is a process in which the kinetic energy of the electrically conductive fluid is transformed into magnetic energy requires  substantial clarification.

Secondly, the magnetic energy growth in 2D is quite similar to that one in 3D. In contrast, the well-known antidynamo theorem \cite{Zeldovich1957} claims  that 2D dynamo is impossible. Again, Zeldovich antidynamo theorem assumes small however finite Ohmic losses and is not relevant for the dynamo stage under consideration. Nevertheless, for this stage the exponential magnetic energy growth was suspected in previous studies of turbulent dynamo \cite{Novikov1983} and now we are able to demonstrate it quantitatively.

It should be also noted that Zeldovich antidynamo theorem supposes that the flow as well as magnetic field are localized in space and have finite total energy, while we consider spatially homogeneous situation. The difference between these two statements of problem was investigated in \cite{Kolokolov}.

The presented results are applicable at the initial stage of dynamo action only and do not allow a straightforward extrapolation to the later stages of dynamo action. In order to
illustrate the specific conditions
of the initial stage  consider the example given in \cite{Z1984}. 

A smooth velocity field $\bf v$ allows a linear approximation nearby a given point. Consider as a particular example the following velocity field:
\begin{equation}
 v_x = cx,\quad \quad v_y=-cy, \quad \quad v_z=0  
 \label{linear}
\end{equation}
with $c>0$ and seed magnetic field $H_x=H_0$, $H_y=H_z=0$. A straightforward calculation shows that $H_x$ grows with growth rate $c$. Note that this formal result also does not contradict the  antidynamo theorem \cite{Zeldovich1957}. The formal reason is that in \cite{Zeldovich1957} the flow and seed magnetic field have finite energy, while here both energies formally diverge. Moreover, \cite{Zeldovich1957} assumes small but nonzero magnetic diffusivity that is neglected here.

Physically, the magnetic field growth obtained in the above example can not continue infinitely. The point is that magnetic loops of any real 2D magnetic field are closed and somewhere there is an oppositely directed part of the magnetic loop. Sooner or later this part will
meet the part under consideration and magnetic field cancellation will stop the magnetic field growth. It is the most important limitation for the first stage of dynamo. 

At the later stages of dynamo action, however, other fundamental effects become important and start to contribute to magnetic field growth. For example, mirror asymmetry of the flow helps to avoid magnetic field cancellation and allows mean magnetic field growth.

The above example demonstrates that straightforward extrapolation of the results obtained under the specific conditions of the initial stage to the later stages of dynamo action can be misleading. 

One more limitation of the model is the consideration of the
Lagrangian approach only. Extrapolation of the results
to the Eulerian approach is not trivial and we leave
it outside the scope of this work. We refer to \cite{Kleeorin2002} for more details.

Our feeling, however, is that the method of extended matrices is a reasonable generalization of Kasantsev \cite{Kazantsev1968} method for
investigation of finite memory effects in turbulent dynamos. In order to perform this generalization one has to consider magnetic field transport for a small yet finite magnetic diffusion, e.g. by considering the random trajectories, which include Wiener diffusion process added to the the conventional Lagrangian trajectories following an approach elaborated in  \cite{Kleeorin2002}. 
To reveal a link between the finite- and short-correlated approach we provide an Appendix B, where the limiting case is considered.

Another possible application of the proposed approach is the study of passive scalar transport in the linear velocity field \cite{Elperin2001, Elperin2002}.

\appendix
\section{Isotropic model}

In this appendix we demonstrate that the method
of energy growth rate estimation, developed in \cite{SI15}
for the isotropic model, results in the same estimates
as we obtain using the method of extended matrix.
It should be noted that once the isotropic condition
is not satisfied, the method proposed in \cite{SI15}
becomes inapplicable. 

The isotropy assumption implies that the
distribution of $\exp (\hat A \tau ){\bf w}$ does not depend on the unit vector $\bf w$.
Then the growth rate  of the $p$-th statistical moment can be expressed in a finite form (\cite{SI15}):
\begin{equation}
    \gamma_p = \frac{1}{2p\tau}\ln\langle\|\exp(\hat{A}\tau)\textbf{w}\|^p \rangle \, ,
\end{equation}
where $\textbf{w}$ is an arbitrary unit vector.
Let us demonstrate that for small $\tau$  
approximation of the matrix exponent $\exp(\hat A\tau)$
with a Taylor series leads to the same estimation
of $\gamma_2$ as in the method of extended matrices.

\subsection{Growth rates for the isotropic 3D case}

For small $\tau$ we obtain an approximation
in terms of the Strouhal number
$s=\tau v/l$:
\begin{equation}
    \langle\|\exp(\hat A\tau)\textbf{w}\|^2\rangle = 1 + 2s^2 + \frac{9}{10}s^4 + \frac{149}{500}s^6 + O(s^8) \, .
\label{vector_norm_3d}
\end{equation}
The vector $\textbf{w}$ can be eliminated from the right hand side
of the equation due to the rotation-invariant distribution of matrix $\hat A$ and as the vector $\textbf{w}$ is a unit vector, we arrive at
(\ref{roots}).

\subsection{Growth rates for the isotropic 2D case}

More substantial simplifications can be obtained for the 2D case, where the matrix exponential $\exp\hat{A}$ can be obtained explicitly.
Let again $\varphi=-\hat A_{11}^2 - \hat A_{12}\hat A_{21} = \det\hat{A}$ (recall that $\hat A_{11} = -\hat A_{22}$), then
\begin{equation}
\exp\hat{A} = \begin{pmatrix}
\cos\sqrt\varphi + \frac{
\hat A_{11}}{\sqrt\varphi}\sin\sqrt\varphi &
\frac{\hat A_{12}}{\sqrt\varphi}\sin\sqrt\varphi \\
\frac{\hat A_{21}}{\sqrt\varphi}\sin\sqrt\varphi & \cos\sqrt\varphi -
\frac{\hat A_{11}}{\sqrt\varphi}\sin\sqrt\varphi
\end{pmatrix} \, .
\label{expB2}
\end{equation}

By averaging $\|\exp\hat{A}\textbf{w} \|^2$ and noting the
symmetry of distributions of $\hat A_{ij}$, along with the facts that $\langle (\hat A_{12}^2/\varphi)\sin^2\sqrt\varphi\rangle  = \langle (\hat A_{21}^2/\varphi)\sin^2\sqrt\varphi\rangle $ and $\|{\bf w}\|=1$, we obtain
\begin{equation}
\lambda = \langle \|\exp\hat{A}\textbf{w}\|^2\rangle  = 
\langle \cos^2\sqrt\varphi + \frac{\hat A_{11}^2 + \hat A_{12}^2}{\varphi}\sin^2\sqrt\varphi\rangle \,. 
\label{lambda_exact}
\end{equation}
This is the same result as  in (\ref{lambda_exp}).

\section{Short-correlated approximation}

A natural comparison for the finite memory effects can be done using the corresponding results for the short-correlated  (also known as delta-correlated) approximation that are summarized in this appendix.
While exponential growth of magnetic energy in 2D at the first stage of dynamo action was already mentioned in \cite{Novikov1983}, here we obtain this result in a rather straightforward way. Of course, magnetic energy decays at later stages of dynamo action when Ohmic losses become important. 

Consider the extended equation for $Z_{ij} = H_iH_j$. Vectors ${\bf Z}(t+\tau)$ and ${\bf Z}(t)$ are linked
via the fundamental matrix:
\begin{equation}
    {\bf Z}(t+\tau) = \exp(\doublehat{A}\tau){\bf Z}(t) =
    (I +\doublehat{A}\tau + \frac{1}{2}\doublehat{A}^2\tau^2 + ...){\bf Z}(t) \,.
\end{equation}
Rewriting and averaging we obtain:
\begin{equation}
    \frac{\langle {\bf Z}\rangle(t+\tau) - \langle {\bf Z}\rangle(t)}{\tau} = 
   \frac{1}{2}\langle \doublehat{A}^2 \rangle\tau 
   \langle {\bf Z}\rangle(t) + o(\tau) \,.
\end{equation}

Now we assume that $\tau v / l$ has finite limit $s$
as $\tau \to 0$, while $\tau^k v^k / l^k \to 0$ for $k>1$. In 2D case we obtain:
\begin{equation}
    \frac{d}{dt}\langle {\bf Z} \rangle = 
    \frac{vs}{2l}\begin{pmatrix}
    1 & 0 & 0 & 3\\
    0 & -1 & -1 & 0\\
    0 & -1 & -1 & 0\\
    3 & 0 & 0 & 1
    \end{pmatrix}\langle {\bf Z} \rangle \,.
\label{z2d}
\end{equation}
From Eq.~(\ref{z2d}) we can separate a closed equation
for the first and the last components of the vector $\langle {\bf Z} \rangle$. Denoting ${\bf H}^2 = (H_1^2,  H_2^2) = (Z_{11}, Z_{22})$ we obtain
\begin{equation}
    \frac{d}{dt}\langle {\bf H}^2 \rangle = 
    \frac{vs}{2l}\begin{pmatrix}
    1  & 3\\
    3 & 1
    \end{pmatrix}\langle {\bf H}^2 \rangle \,.
\end{equation}
This results in the following 
equation for mean energy $\langle \| {\bf H} \|^2 \rangle$:
\begin{equation}
    \frac{d}{dt}\langle \| {\bf H} \|^2 \rangle = 
    \frac{2vs}{l}\langle \| {\bf H} \|^2 \rangle \,.
\label{H2_dcorr_eq}
\end{equation}
The (non-normalized) growth rate is obviously $2vs/l$.

Similarly, in 3D case denoting  ${\bf H}^2 = (H_1^2,  H_2^2,  H_3^2)$ we obtain:
\begin{equation}
    \frac{d}{dt}\langle {\bf H}^2 \rangle = 
    \frac{2vs}{5l}\begin{pmatrix}
    1 & 2 & 2\\
    2 & 1 & 2\\
    2 & 2 & 1
    \end{pmatrix}\langle {\bf H}^2 \rangle \,.
\end{equation}
As a consequence, mean energy $\langle \| {\bf H} \|^2 \rangle$ follows the equation
\begin{equation}
    \frac{d}{dt}\langle \| {\bf H} \|^2 \rangle = 
    \frac{2vs}{l}\langle \| {\bf H} \|^2 \rangle \,.
\end{equation}
The non-normalized growth rate is obviously $\lambda=2vs/l$.

Appropriate normalization for the second moment (the factor 1/4) then results in the following normalized growth rate (omitting $v/l$)
\begin{equation}
   \gamma_2 = \frac{1}{2}s \, ,
\end{equation}
which is exactly the same result as we got for finite $\tau$.

Note that if we additionally assume $\tau^kv^k/l^k\to s^k$ for e.g. $k=3$ and 5, and $\tau^kv^k/l^k\to 0$ for $k>5$,
we obtain instead of (\ref{H2_dcorr_eq}) an extended equation given by
\begin{equation}
    \frac{d}{dt}\langle \| {\bf H} \|^2 \rangle = 
    \frac{v}{l}\left(2s + \frac{2}{3}s^3 + \frac{4}{9}s^5\right)\langle \| {\bf H} \|^2 \rangle \, ,
\end{equation}
which is analogous to (\ref{lambda_iso_2d}).
Now let's assume $\tau^k v^k/l^k\to s^k$ for all $k$. 
Also let's rewrite ${\lambda}(s)$ in (\ref{H2equation}) as ${\lambda}(s) = 1 + \omega(s)$ to obtain
\begin{equation}
    \frac{d}{dt}\langle \| {\bf H} \|^2 \rangle = 
    \frac{v}{l}\frac{\omega(s)}{s}\langle \| {\bf H} \|^2 \rangle \,.
\label{v1}
\end{equation}
For comparison, an equation (\ref{H2equation}) implies
\begin{equation}
    \frac{\langle \| {\bf H} \|^2\rangle(t+\tau) - \langle \| {\bf H} \|^2\rangle(t)}{\tau} = 
    \frac{\omega(v\tau/l)}{\tau}\langle \| {\bf H} \|^2\rangle(t) \,.
\label{v2}
\end{equation}
We find that (\ref{v1}) is a limit of 
(\ref{v2}) as $\tau\to 0$ if
\begin{equation}
   \lim\limits_{\tau\to 0} \frac{\omega(v\tau/l)}{v\tau/l} = 
   \frac{\omega(s)}{s} \,.
\label{}
\end{equation}

Similarly, we find that the growth rate according to 
(\ref{v1}) (that is $\omega(s)/s[v/l]$) is a limit of the
the growth rate according to 
(\ref{v2}) (that is $(1/\tau)\ln(1+\omega(v\tau/l))$) as $\tau\to 0$
if
\begin{equation}
   \lim\limits_{\tau\to 0} \frac{1}{v\tau/l}\ln(1+\omega(v\tau/l)) = 
   \frac{\omega(s)}{s} \,,
\label{}
\end{equation}
{or, denoting $\delta=v\tau/l$, we obtain}
\begin{equation}
   \lim\limits_{\delta\to s} \frac{1}{\delta}\ln(1+\omega(\delta)) = 
   \frac{\omega(s)}{s} \,.
\label{}
\end{equation}

In the axisymmetric 2D model for $u\sim 1$, $L\sim 1$ the short-correlated approximation is
\begin{equation}
    \frac{d}{dt}\langle {\bf H}^2 \rangle = 
    \frac{\sqrt{2}v_1s}{2l_1}\begin{pmatrix}
    1 & u+2L\\
    6-3L & 1\\
    \end{pmatrix}\langle {\bf H}^2 \rangle \,.
\end{equation}

Up to a scale factor of $\sqrt{2}v_1/l_1$, approximation of the leading eigenvalue is
\begin{equation}
    \lambda_1 = 2s\left(1+\frac{u-L}{8}\right) + O(s^3, u^2, L^2) \,.
\end{equation}
Corresponding normalized growth rate is $\gamma_2=(1/4)\lambda_1$.

In the axisymmetric 3D case we have
\begin{equation}
    \frac{d}{dt}\langle {\bf H}^2 \rangle = 
    \alpha\begin{pmatrix}
    1 & u+L & u+L\\
    4-2L & 1+u-L & 2+3(u-L)\\
    4-2L & 2+3(u-L) & 1+u-L
    \end{pmatrix}\langle {\bf H}^2 \rangle \,,
\end{equation}
where $\alpha=2\sqrt{3}v_1s/5l_1$.

Up to a scale factor of $\sqrt{3}v_1/l_1$, approximation of the leading eigenvalue is given by
\begin{equation}
    \lambda_1 = 2s\left(1+\frac{2(u-L)}{3}\right) + O(s^3, u^2, L^2) \,.
\end{equation}
and the corresponding normalized growth rate is given by $\gamma_2=(1/4)\lambda_1$.

\begin{acknowledgments}
We thank the reviewers for valuable comments and suggestions.
We thank Naga Varun for critical reading of the manuscript.
EI acknowledges the partial support of RSF grant 20-72-00106 for numerical simulations.
DS acknowledges the support of the Russian Ministry of Science and Higher Education, agreement No. 075-15-2019-1621.
\end{acknowledgments}

\bibliography{apssamp}

\end{document}